\documentclass[12pt]{iopart}
\usepackage{graphicx}
\usepackage{subfig}

\begin{document}

\title[An improved algorithm for narrow-band searches of continuous gravitational waves]{An improved algorithm for narrow-band searches of continuous gravitational waves}

\author{S. Mastrogiovanni$^{1,2}$, P. Astone$^{2}$, S. D'Antonio$^{3}$, S. Frasca$^{1,2}$, G. Intini$^{1,2}$, P. Leaci$^{1,2}$, A. Miller$^{1,4}$, C. Palomba$^{2}$, O. J. Piccinni$^{1,2}$, A. Singhal$^{2}$}

\address{$^1$Universita di Roma ’La Sapienza’, I-00185 Roma, Italy}
\address{$^2$INFN, Sezione di Roma, I-00185 Roma, Italy}
\address{$^3$INFN, Sezione di Roma Tor Vergata, I-00133 Roma, Italy}
\address{$^4$University of Florida, Gainesville, FL 32611, USA}

\ead{simone.mastrogiovanni@roma1.infn.it}
\vspace{10pt}
%\begin{indented}
%\item[]February 2014
%\end{indented}

\begin{abstract}
Continuous gravitational waves signals, emitted by asymmetric spinning neutron stars, are among the main targets of current detectors like Advanced LIGO and Virgo. In the case of sources, 
like pulsars, which rotational parameters are measured through electromagnetic observations,  
typical searches assume that the gravitational wave frequency is at a given known fixed ratio with respect to the star rotational frequency. 
For instance, for a neutron star rotating around one of its principal axis of inertia the gravitational signal frequency would be exactly two times the rotational frequency of the star. 
It is possible, however, that this assumption is wrong. This is why search algorithms able to take into account a possible small mismatch between the gravitational waves frequency and the 
frequency inferred from electromagnetic observations have been developed. In this paper we present an improved pipeline to perform such narrow-band searches for continuous gravitational 
waves from neutron stars, about three orders of magnitude faster than previous implementations. 
The algorithm that we have developed is based on the {\it 5-vectors} framework and is able to perform a fully coherent search over a frequency band of width $\mathcal{O}$(Hertz) and for hundreds of spin-down values running a few hours on a standard workstation. This new algorithm opens the possibility of long coherence time searches for objects which rotational parameters are highly uncertain.
\end{abstract}

% Uncomment for PACS numbers
%\pacs{00.00, 20.00, 42.10}
%
\vspace{2pc}
\noindent{\it Keywords\/}:Neutron star, Gravitational waves, interferometric detectors
%
% Uncomment for Submitted to journal title message
%\submitto{\JPA}
%
% Uncomment if a separate title page is required
%\maketitle
% 
% For two-column output uncomment the next line and choose [10pt] rather than [12pt] in the \documentclass declaration
%\ioptwocol
%
\section{Introduction}
On 14$^{th}$ September 2015 the analysis, by the LIGO and Virgo Collaborations, of the data collected by the two LIGO interferometers during the science run O1, allowed to detect for the first time the gravitational waves (GW) emitted by a binary black hole merger \cite{abbot:detection}, followed on 26$^{th}$ December 2015 by the detection of  a second event again associated to a binary black hole merger\cite{abbot:detection2}. These two events have started the era of gravitational waves astronomy. In fact, other sources of GWs are expected to emit signals potentially detectable by current interferometers. Among these there are spinning neutron stars (NS) asymmetric with respect to the rotation axis, which are expected to emit continuous, nearly monochromatic signals (CW). In the standard case of an asymmetric star rotating around one of its principal axis of inertia, the signal frequency is twice the rotational frequency of the star. 
The expected signal amplitude is very small but one can exploit the signal long duration, at least in principle, to build-up the signal-to-noise ratio to detectable levels.  

Depending on the degree of knowledge of the NS parameters, different types of searches can be done. \textit{Targeted searches} for CW from known pulsars are the most sensitive 
\cite{jasi:gw_known_pulsars} \cite{matt:bayes} \cite{krolak:fstat}. They are based on the application of matched filtering and require a very accurate knowledge of the source parameters: 
right ascension $\alpha$, the declination $\delta$ and the rotational frequency of the NS $f_{rot}$ with its derivatives $\dot{f}_{rot}$, $\ddot{f}_{rot},...$. The knowledge of the source 
parameters 
comes from electromagnetic observations at different wavelengths. On the other extreme, \textit{blind searches}, which are based on hierarchical approaches, are able to explore a wide volume in the parameter space at the cost of a lower sensitivity \cite{astone:allsky}\cite{powerflux}\cite{fstat}\cite{skyhough}\cite{einathome}.

\textit{Narrow-band searches} are an extension of targeted searches, in which the position of the source is assumed to be accurately known while the rotational parameters are slightly uncertain. This type of search can still be based on matched filtering but, of course, is computationally heavier with respect to targeted searches. In general, narrow-band searches allow to take into account a possible mismatch between the GW rotational parameters and those inferred from electromagnetic observations. For instance, the GW signal could be emitted by the core of the NS which may have a slightly different rotational frequency with respect to the magnetosphere. Secondly, in order to make a targeted search for a given object updated ephemeris covering the time span of the data at hand are needed. If they are not available, a fully coherent search based on wrong ephemeris could introduce a phase error with a consequent loss of signal-to-noise ratio \cite{ashton:snr_logg}. In recent years an analysis pipeline able to make a fully coherent search over a fraction of Hertz and tens of spin-down values has been developed \cite{rob:method} and applied to interferometric detector data \cite{rob:obs}.

In this paper we present an improved, and computationally cheap, algorithm for narrow-band searches which is also suited to coherently explore a wider volume
of the parameter space. The paper is organized as follows. In section \ref{sec:II} we summarize the main features of CW signals. In section \ref{sec:III} we
describe the new narrow-band pipeline, stressing the improvements with respect to the previous implementation. In section \ref{sec:IV} we describe the
validation tests of the algorithm, done using both software and hardware simulated CW signals. In this section we also discuss the computational load of the pipeline. In section \ref{sec:V} we present a case study discussing the search for CW from the central compact object G353.6-0.7 in Virgo detector VSR4 data. Finally, in Sec. \ref{sec:VI} we discuss the future perspectives for the application of this new pipeline.

\section{The signal}
\label{sec:II}
The CW signal emitted by an asymmetric spinning neutron star rotating around a principal axis of inertia can be written, following the formalism first introduced in \cite{pia:articolo}, 
as:
\begin{equation}
h(t)= H_0 ( H^+ A_+ (t) + H^\times A_\times (t)) e^{2 \pi i f_{gw} (t)}
\label{eq:Hgrande}
\end{equation}
where taking the real part is understood.
The complex amplitudes $H^+, H^\times$ are given by:
\begin{eqnarray*}
H^+ =\frac{\cos(2 \psi) - i \eta \sin (2 \psi)}{\sqrt{1+\eta^2}} \qquad H^\times =\frac{\sin(2 \psi) - i \eta \cos (2 \psi)}{\sqrt{1+\eta^2}} \\ 
\end{eqnarray*}
where $\eta$ is the ratio of the polarization ellipse semi-minor to semi-major axis and the polarization angle $\psi$ is defined as the direction of the major axis with respect to the celestial parallel of the source (measured counter-clockwise).
The parameter $\eta$ takes the values from -1 to 1 (in particular,  $\eta=\pm 1$  if the wave is circularly polarized clockwise or counter-clockwise and $\eta=0$ if the wave is linearly polarized).
The functions $A_+ (t),\, A_\times (t)$ are the detector \textit{sidereal responses} and encode the interferometer response to the GW. The signal at 
the detector is not monochromatic, i.e. the frequecy $f_{gw}=2f_{rot}$ in Eq. (\ref{eq:Hgrande}) is not constant. In fact it is modulated by some effects such as the \textit{Romer Delay}, 
due to 
the detector motion, and the source intrinsic spin-down, due to energy loss from the source. All these effects must be properly taken into account in order to increase the signal to noise ratio in the analysis.
It can be shown that the waveform defined by Eq. \ref{eq:Hgrande} is equivalent to the GW signal expressed in the more standard formalism of \cite{jasi:gw_known_pulsars}, given the following relations: 
\begin{equation}
\eta=-\frac{2\cos \iota}{1+\cos^2 \iota},
\label{eq:etaiota}
\end{equation}
where $\iota$ is the angle between the line of sight to the source and the star rotation axis, and
\begin{equation}
H_0=h_0 \sqrt{ \frac{1+6 \cos ^2 \iota + \cos^4 \iota}{4}} 
\end{equation}
%\begin{equation}\fl
%h(t)=\frac{1}{2}F_+(t,\psi) h_0 (1+ \cos^2 \iota) \cos  2 \pi f_{GW}(t)+  F_\times h_0 \cos \iota \sin ( 2 \pi f_{GW}(t))
%\label{eq:LIGO_h}
%\end{equation}
 with
\begin{equation}
h_0=\frac{1}{r} \frac{4 \pi^2 G }{c^4} I_{zz} f_{gw}^2 \epsilon
\label{eq:GW_amplitude}
\end{equation}
being $d,I_{zz}$ and $\epsilon$ respectively the distance, the NS's moment of inertia with respect to the rotation axis and the {\it ellipticity}, which measures the star degree of 
asymmetry.

Given a source with measured rotation frequency $f_{rot}$, frequency derivative $\dot{f}_{rot}$ and distance $d$, the GW signal amplitude can be constrained assuming that all the 
rotational energy is lost via gravitational radiation. This absolute upper limit, called {\it spin-down limit}, is given by: 
\begin{equation}
h_{sd}=8.06 \cdot 10^{-19} I_{38} \bigg(\frac{1kpc}{d} \bigg) \bigg(\frac{\dot{f}_{rot}}{Hz/s} \bigg)^{1/2} \bigg(\frac{Hz}{f_{rot}} \bigg)^{1/2}
\label{eq:sd_limit}
\end{equation} 
being $I_{38}$ the star moment of inertia in units of $10^{38} kg \, m^2$. The corresponding spin-down limit on the star "fiducial" ellipticity
\cite{jasi:gw_known_pulsars} is
\begin{equation}
\epsilon_{sd}=0.307 \bigg( \frac{h_{sd}}{10^{-24}}\bigg) f_{rot}^{-2} I_{38}^{-1} \bigg(\frac{d}{1 kpc} \bigg)
\end{equation} 
Eq. \ref{eq:sd_limit} can be re-written using the age parameter $\tau=\frac{f_{rot}}{\dot{f}_{rot}}$ and assuming that the NS was born with a spin much higher than the one that we observe 
now:
\begin{equation}
h_{sd,\tau} \simeq 1.24 \times 10^{-24} \left( \frac{3.3 kpc}{d} \right) \sqrt{\left( \frac{I_{zz}}{10^{45} g cm^2} \right) \left( \frac{300 yr}{\tau} \right) }
\label{eq:sdlimit}
\end{equation}
For a given source, even in absence of a detection, establishing an upper limit below the spin-down limit is an important milestone, as it allow us to put a non-trivial 
constraint on the fraction of rotational energy lost through GWs.

\section{Narrow-band search}
\label{sec:III}
In this section we remind the main general features of a narrow-band search, and discuss what are the main improvements we have introduced with respect to the original pipeline firstly described in \cite{rob:method}. In a narrow-band search the source position in the sky is assumed to be known and the analysis is run over a grid built in the frequency/spin-down space. The number of points in the grid obviously depends on the range of frequency, $\Delta f$, and spin-down, $\Delta \dot{f}$, that we want to explore, and on the grid steps which, for a fully coherent analysis, depend only on the total observation time $T_{obs}$. More precisely, the number of frequency bins is given by the ratio between the frequency range $\Delta f$ and the frequency bin width $\delta f=\frac{1}{T_{obs}}$:
\begin{equation}
n_f = \Delta f T_{obs} \approx  3 \cdot 10^7 \bigg( \frac{\Delta f}{1 Hz} \bigg) \bigg(  \frac{T_{obs}}{1 yr}\bigg),
\end{equation}
while the number of spin-down bins is given by the ratio of the spin-down range $\Delta \dot{f}$ and the spin-down bin width $ \delta \dot{f}=1/ T^2 _{obs}$\footnote{The spin-down bin is defined as that spin-down value such that during the observation time $T_{obs}$ the frequency variation due to it is within a frequency bin.} :
\begin{equation}
 n_{\dot{f}} = \Delta \dot{f} T^{2}_{obs} \approx 900 \bigg( \frac{\Delta \dot{f}}{10^{-12} Hz} \bigg) \bigg(  \frac{T_{obs}}{1 yr}\bigg)^2 
\end{equation}
Similarly, the number of values for the second order spin-down term is given by
\begin{equation}
 n_{\ddot{f}} = \Delta \ddot{f} T^{3}_{obs}/2 \approx 0.16 \bigg( \frac{\Delta \ddot{f}}{10^{-23} Hz} \bigg) \bigg(  \frac{T_{obs}}{1 yr}\bigg)^3 
\end{equation}
For reasonably small ranges of the second order spin-down parameter $ n_{\ddot{f}}=1$, then the total number of points explored in the parameter space is then  $n_{tot}=n_f n_{\dot{f}}$, which scales as $T_{obs}^3$.

\begin{figure}[h!]
\hspace*{1.7cm}
\includegraphics[scale=0.35]{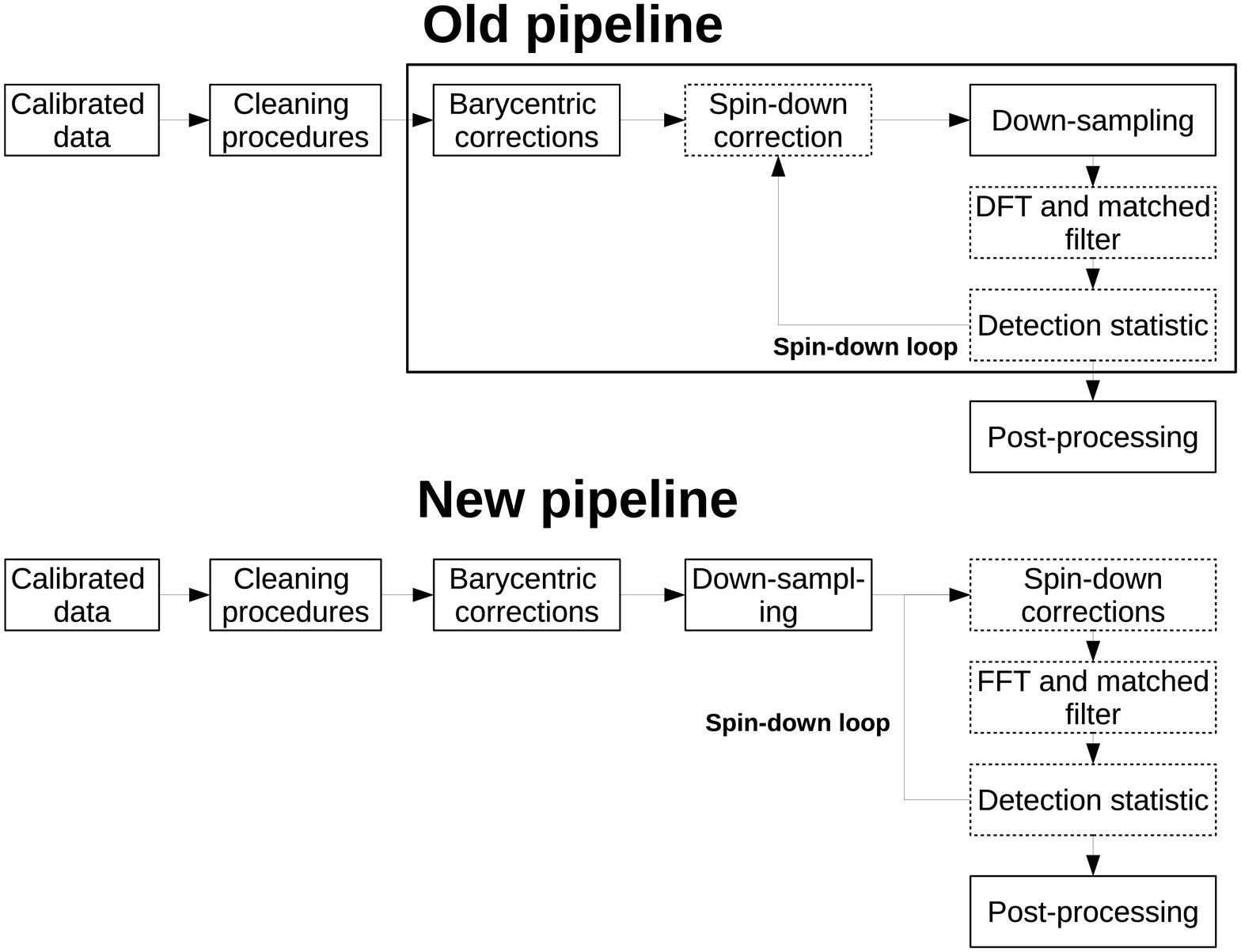}
\caption{Top Panel: Flowchart of the old pipeline  In particular the procedures that have been improved are indicated in a dashed box. More details on the pipeline  can be found in \cite{rob:method}. Bottom Panel: Flowchart of the new pipeline, the procedures indicated in the dasehed box will be described in this paper.}
\label{fig:flowchart}
\end{figure}

The flow charts of the old and the new pipelines are shown if Fig. \ref{fig:flowchart}. Some of the core parts of the analysis, highlighted by dashed boxes, have a different implementation in the two cases. Let us start briefly describing what are the main steps of the old  pipeline.  

For a given detector and source position, the barycentric corrections are applied by introducing a new time variable defined as
\begin{equation}
t'= t+ \Delta_{R} + \Delta_{E} -\Delta_{S}
\label{eq:tprimo}
\end{equation}
where $\Delta_{R},\, \Delta_{E},\, \Delta_{S}$ are, respectively, the Romer, Einstein and Shapiro delay. Seen in the new time variable $t'$, which is the time
in the Solar System Barycenter, a signal in the data would not be subject to the barycentric modulation anymore. It is important to stress that the time $t'$
does not depend on the frequency, hence the barycentric corrections are computed just once and hold for the whole frequency band. The spin-down correction is
done using the same technique, re-defining the time as:
\begin{equation}
\tau= t'+ \frac{\dot{f}}{2f}(t'-t_0')+ \frac{\dot{f}}{2f}(t'-t_0')^2
\end{equation}
It is clear that the time $\tau$ depends explicitly on the frequency and spin-down values.
Seen in the non-uniformly sampled time variable $\tau$, the signal is now monochromatic apart from the amplitude and phase modulation encoded in the sidereal responses.
Once the corrections have been done the data are down-sampled at 1 Hz. Now, for each frequency and spin-down bin of the grid, a set of five discrete Fourier Transforms (DFT) is evaluated in order to get the so-called {\it 5-vectors}. Given a time series, a 5-vector consists of the Fourier components at the five frequencies at which the power of a CW signal is split by the sidereal modulation:
\begin{equation}
G_k(f)=\sum_{i=0}^{N_{s}} s(t_i)e^{- 2 \pi i(f-k F_{\oplus})t_i}dt
\end{equation}
where $F_{\oplus}$ is the Earth's sidereal frequency and $k_i=-2,-1,0,1,2$. See \cite{pia:articolo} for more details. Then, two matched filters are computed in the frequency domain by making the scalar product between the data 5-vector $\vec{X}$ and the  normalized sidereal response 5-vectors $\vec{A}_{+ / \times} / |\vec{A}_{+ / \times}|^2 $: 
\begin{equation}
 \widehat{H}^{+/ \times} (f) =\vec{X} (f) \cdot \frac{\vec{A}_{+ / \times}}{|\vec{A}_{+ / \times}|^2} 
\end{equation}
where the "$\cdot$" implies taking the complex conjugate of the second term.
The quantities $\widehat{H}^{+/ \times}$ are estimators of the CW complex amplitudes, defined in Eq.\ref{eq:Hgrande}, and are used to build the detection statistic: 
\begin{equation}
S=|\vec{A}_{+}|^4 |\widehat{H}^+|^2 +|\vec{A}_\times|^4 |\widehat{H}^\times|^2
\end{equation}
Once we have obtained a value of the detection statistic for each frequency and spin-down bin, we select the most interesting candidates for the follow-up. Typically, this is done by setting a threshold on the detection statistic, corresponding to a small p-value on the noise-only distribution (e.g. 1\%, after taking into account the number of trials), and taking the candidates with a detection statistic above the threshold \cite{rob:obs}.

The computational cost of the pipeline described so far is dominated by the computation of a large number of DFTs, one for each frequency and spin-down bin. 
\subsection{Main features of the new pipeline}
In the new pipeline, the barycentric corrections are done exactly in the same way as in the old pipeline.
Let us now describe which are the main improvements in the rest of the pipeline, which allow to dramatically reduce the computational load of the analysis and, at the same time, also improve the sensitivity. They can be summarised as: \textit{a)} application of the spin-down correction in phase and after the down-sampling ; \textit{b)} use of the Fast Fourier Transform algorithm (FFT) to compute 5-vectors over the frequency grid; \textit{c)}  FFT interpolation at half-bins in order to reduce the sensitivity loss due to the discretization. 
\subsubsection{Spin-down correction}
The spin-down correction is now applied in phase rather than in time. The frequency evolution of the signal emitted by a spinning NS can be expressed using a Taylor expansion of the frequency with respect to time:
\begin{equation}
f(t)= f(t_0)+  \sum_{k\ge } \frac{1}{k!} \frac{d^k f}{dt^k} \bigg{|}_{t=t_0} (t-t_0)^k
\label{eq:sd}
\end{equation}
Theoretically, we can have any value for the order $k$ but, practically, for most known pulsars only the first 2-3 terms are measurable.
The corresponding phase evolution is given by
\begin{equation}
\phi_{sd} (t)= 2 \pi \int_{t_0}^{t} \sum_{k\ge 1} \frac{1}{k!} \frac{d^k f}{dt^k} \bigg{|}_{t=t_0} (t-t_0)^k dt =  2 \pi \sum_{k\ge 1} \frac{1}{(k+1)!} \frac{d^k f}{dt^k} \bigg{|}_{t=t_0} (t-t_0)^{k+1}
\label{eq:sd_correction}
\end{equation}
The frequency variation, due to the spin-down of a possible signal into the data, can be removed by multiplying the data by the factor $e^{-i 2 \pi  \phi_{sd} (t)} $. In this way the signal would become 
\begin{equation}
h'(t)=h(t)\cdot e^{- 2 \pi i \phi_{sd}}=H_0(H^{+}A_{+}(t)+H^{\times}A_{\times}(t))e^{i (f_{gw} t + \phi_0)}
\end{equation} 
which is monochromatic apart from the amplitude and phase modulation due to the sidereal responses. T0he spin-down corrections are applied after the down-sampling, thus reducing the computational cost of the algorithm. 
%Most importantly, the phase factor of Eq. \ref{eq:sd_correction} does not depend on the frequency. 
%This means that for a given spin-down bin we can apply this corrections only once.

\subsubsection{Replacing the Discrete Fourier Trasform with the Fast Fourier Trasform }
As shown in previous sections, we previously computed the DFT in every frequency bin in order to identify the 5-vectors. The computational cost of the DFT is the main bottleneck if it must be performed a large number of times.
To overcome this problem we have modified the original pipeline (written in Matlab) replacing the computation of the DFT for each frequency bin with an FFT.
The FFT is computed in Matlab using the Turkey-Cooley algorithm \cite{turkey:FFT}. Before proceeding, let us see how the DFT and the FFT are related. The value
of the DFT of a time series with samples $x_j$, at the frequency $f_k$, is given by:
\begin{equation}
\tilde{x}_{DFT}(f_k)=\sum_{j=0}^{N-1} x_{j} e^{-2 \pi i j \delta t f_k} \delta t
\end{equation}
where $i$ the imaginary unit  and $ \delta t $ is the sampling time. With some algebraic manipulations we find
\begin{equation}
e^{-2 \pi i \delta t f_k}\tilde{x}_{DFT}(f_k)=\sum_{j=1}^{N} x_{j-1} e^{-2 \pi i  j  \delta t f_k} \delta t
\label{eq:DFT}
\end{equation}
The FFT is defined in Matlab as
\begin{equation}
\tilde{x}_{FFT}(k)=\sum_{j=1}^{N} x_j e^{-\frac{2 \pi i}{N}(j-1)(k-1)}=\sum_{j=1}^{N} x_j e^{-2 \pi i \delta t \delta f(j-1)(k-1)}
\end{equation}
where $\delta f$ is the frequency bin. After further manipulations we obtain
\begin{equation}
\tilde{x}_{FFT}(k)=e^{2 \pi i \delta t \delta f (k-1)} \sum_{j=1}^{N} x_j e^{-2 \pi i \delta t \delta f j (k-1)}
\label{eq:FFT}
\end{equation}
Given that $k \cdot \delta f=f_k$ and $j \cdot \delta t=t_j$ are the time samples, we have:
\begin{eqnarray*}
& \tilde{x}_{FFT}(k=1)=\tilde{x}_{FFT}(f=0)=\sum_{j=1}^{N} x_j \\
&\tilde{x}_{FFT}(k=2)=\tilde{x}_{FFT}(f=\delta f)=e^{2 \pi i \delta t \delta f}  \sum_{j=1}^{N} x_j e^{-2 \pi i  t_i \delta f } 
\end{eqnarray*}
and so on. By comparing Eq. \ref{eq:DFT} and  Eq. \ref{eq:FFT} we obtain the simple relation:
\begin{equation}
\tilde{x}_{DFT} (f)=\tilde{x}_{FFT} (f) \cdot \delta t
\end{equation}
Hence, the following relations for the complex amplitude estimators follow: 
\begin{equation}
\widehat{H}^{+/ \times}_{FFT}(f)=\widehat{H}^{+/ \times}_{DFT}(f)
\end{equation}
while for the detection statistic we have
\begin{equation}
S_{FFT}(f_k)\cdot \delta t^4=S_{DFT}(f_k)
\label{eq:rel_dt}
\end{equation}
In order to fully exploit the use of the FFT in the computation of 5-vectors, a suitable grid in frequency must be built. More precisely, the sidereal angular frequency of the Earth is chosen to be an integer multiple of the frequency bin. In this  way the 5-vector components correspond to integer frequency bins of the FFT. This is done by properly adjusting the observation time in such a way that the frequency bin is a sub-multiple of the sidereal angular frequency. Typically, this implies cutting away a few hours of data with a completely negligible effect on the search sensitivity. 
The use of the FFT, together with this adjusted frequency grid, allows to compute the entire ensemble of 5-vectors over a given frequency range with only one application of the FFT algorithm, thus providing a huge gain in computational speed.

\begin{figure}[h]
\hspace*{-0.8cm}
\includegraphics[scale=0.32]{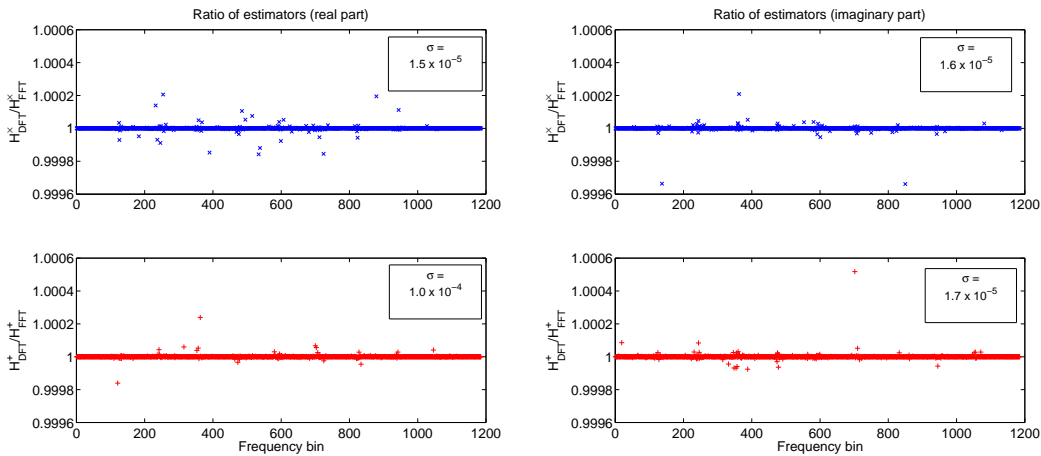}
\caption{The plots show the ratio of the amplitude estimators, computed respectively with the $DFT$ and $FFT$, as a function of the frequency bin on simulated
Gaussian noise data covering 4 months ( for both the real and imaginary part). Top left: ratio of the real part of the $\times$ estimators. Top right: ratio of
the imaginary part of the $\times$ estimators. Bottom left: ratio of the real part of the $+$ estimators. Bottom right: ratio of the imaginary part of the $+$
estimators. The standard deviation is indicated in the boxes.}
\label{fig:H}
\end{figure}

\begin{figure}[h!]
\hspace*{-2.25cm}
\includegraphics[scale=0.38]{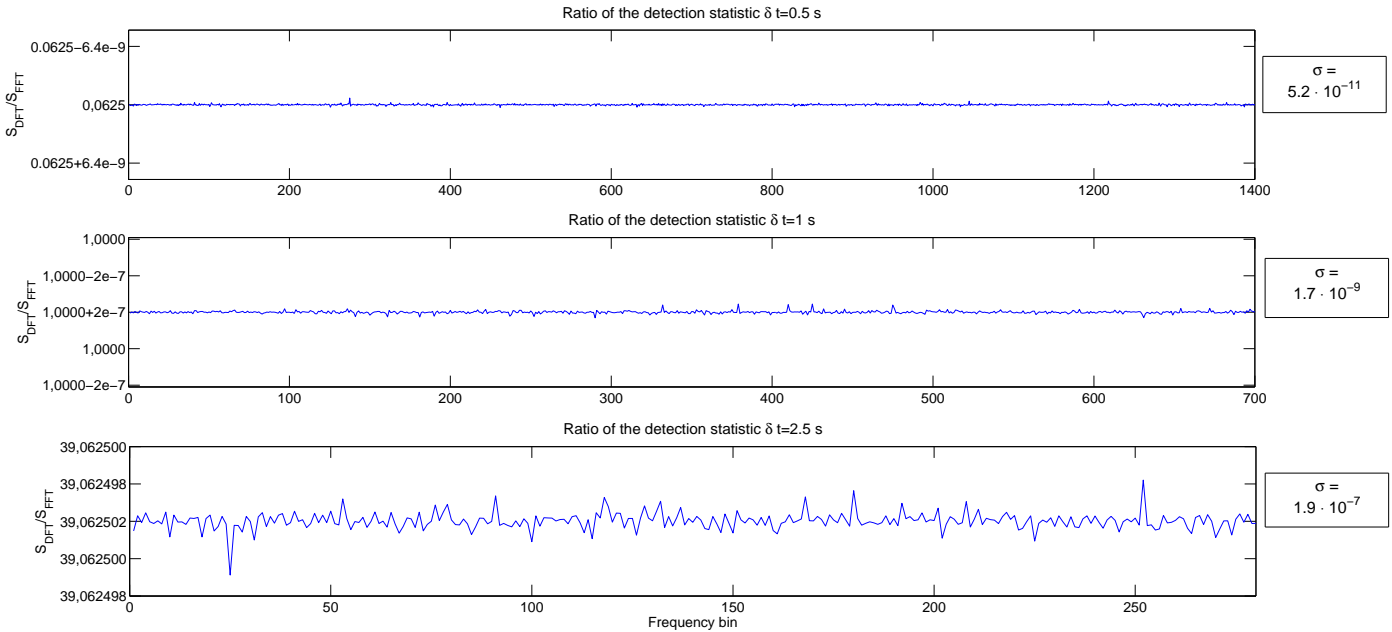}
\caption{Ratio of the detection statistics computed with the DFT and the FFT as a funcition of the frequency bin, using simulated Gaussian data at various sampling times $\delta t$. Top Panel: $\delta t=0.5 s $. Central Panel: $\delta t=1.0 s $. Bottom Panel: $\delta t=2.5 s $. As expected from Eq. \ref{eq:rel_dt}, the variance of the mismatch is a function of the sampling rate, with higher sampling rates providing a smaller difference.}
\label{fig:DS_dt}
\end{figure}

We have done several checks of consistency between the DFT-based and the FFT-based algorithm.
A first check consisted in comparing the estimators $\widehat{H}^{+/ \times}$ computed in the two cases on simulated Gaussian noise. The standard deviation of the mismatch, defined as 
$\frac{Q_{DFT}}{Q_{FFT}}-1$, where $Q$ is the real or imaginary part of $\widehat{H}^{+/ \times}$, is up to a few units per $\sim 10^{-4}$ , see Fig. \ref{fig:H}. 
\clearpage
The next check consisted in comparing the values of the detection statistic obtained using the DFT and the FFT algorithms. We have tested the relation given by Eq. \ref{eq:rel_dt} for sampling rates $\delta t$ different from $1s$ proving that it is valid apart from a relative error smaller than $\sim 10^{-5}$, see Fig. \ref{fig:DS_dt}.
 
\subsubsection{FFT interpolation}
Due to the discretizzation of the frequency, it is possible that the signal's frequency falls between two frequency grid points. This produces a loss of signal-to-noise ratio of up to $ \sim 36 \%$. In order to reduce this loss a standard and cheap method consists in estimating the FFT values at half bin using the ``interbinning'' approximation, see e.g. \cite{ransom:fourier}:
\begin{equation}
\tilde{x}_{FFT,k+\frac{1}{2}} \simeq \frac{1}{4} \pi (\tilde{x}_{FFT,k} - \tilde{x}_{FFT,{k+1}})
\end{equation} 
where the index $k$ denotes the $k$-th frequency bin. We have added this feature to the narrow-band search pipeline. Interbinning has some weak points, like
the fact it introduces a correlation between half and integer bins. An alternative, which avoids correlations but double the time series length, and then the
computational cost, would consist in using zero-padding, adding a number of zeros equal to the original data length. As discussed in \cite{ransom:fourier},
however, interbinning can be usefully used in the detection stage, eventuallys relying on more sophisticate, and computationally heavier, techniques to make deeper searches in the follow-up stage.

\section{Full pipeline tests}
\label{sec:IV}
In this section we describe two kinds of tests done using the new narrow-band search pipeline. The first one consisted in analyzing Virgo VSR4 data searching for a few simulated CW signals directly injected into the interferometer (called {\it hardware injections}). The aim of the test was to verify if we are able to detect these signals and accurately estimate their parameters. The second test consisted in running a narrow-band search for the Crab pulsar again in Virgo VSR4 data. This analysis has been already done, on the same data set, using the old pipeline \cite{rob:obs}. Here we mainly want to make a comparison of the pipelines speed.

\subsection{Hardware injections}
Hardware injections are artificial signal injected during detector observation runs. For the pipeline's test we have chosen three hardware injections, which parameters are shown in Tab. \ref{tabl2}.
\begin{table}[h!]
\begin{center}
\caption{Parameters of the hardware injections used to test the algorithm. $H_0$ is the signal amplitude, $f$ is the signal frequency, $\dot{f}$ is the spin-down, $\alpha$ the right ascension, $\delta$ the declination, $\eta$ the ratio between the polarization ellipse semi-minor and semi-major axes and $\psi$ is the wave polarization angle.}
\label{tabl2}
\hspace*{-1.0cm}
\begin{tabular}{cccccccc}
\hline
HI name&$f~$[Hz]&$\dot{f}~$[Hz/s]&$H_0$&$\eta$&$\psi$[deg]&$\alpha$ [deg]&$\delta$ [deg]\\ \hline
Pulsar2&$575.16357300$&$1.37 \cdot 10^{-13}$&$5.2844 \cdot 10^{-24}$&$0.997$&$77.292$&$215.256166$&$3.443989$ \\
Pulsar3&$108.85715940$&$1.46 \cdot 10^{-17}$&$8.2961 \cdot 10^{-24}$&$0.160$&$25.439$&$178.372574$&$−33.436602$ \\
Pulsar8&$192.23705881$&$8.65 \cdot 10^{-9}$&$8.0673 \cdot 10^{-24}$&$-0.147$&$9.7673$&$351.389582$&$−33.418516$ \\
\hline
\end{tabular}
\end{center}
\end{table}\newline
For the tests we have used the entire VSR4 run of Virgo. The run duration is about $T_{obs}\approx 8.14 \cdot 10^{6} s$. Therefore, we have the following bin width for the frequency, first order spin-down and second order spin-down, respectively:
\begin{eqnarray*}
& \delta f =\frac{1}{T_{obs}} \simeq 1.23 \cdot 10^{-7} \, \textrm{Hz}  \qquad \delta \dot{f} =\frac{1}{T_{obs}^2} \simeq 1.51 \cdot 10^{-14} \, \textrm{Hz}/s \\ & \qquad \qquad  \qquad \delta \ddot{f} =\frac{2}{T_{obs}^3} \simeq 3.70 \cdot 10^{-21} \, \textrm{Hz}/s^2 
\end{eqnarray*}\newline
We have ran the analysis in a small volume of the frequency / spin-down space around the parameters of each injected signal, see Tab \ref{tablII}. In all the
cases, the loudest value of the detection statistic corresponds to the frequency and spin-down bins nearest to the injected parameters.

\Table{\label{tablII} Parameter space explored around three hardware injections. $\Delta f$ is the frequency range and $n_f$ the corresponding number of bins,
$\Delta \dot{f}$ is the spin-down range and $n_{\dot{f}}$ the corresponding number of bins. Finally, $n_{tot}=n_f\cdot n_{\dot f}$ is the total number of points.}
\br
Name&$\Delta f$ [Hz]&$\Delta \dot{f}$ [Hz/s] &$n_f$ &$n_{\dot{f}}$ &$n_{tot}$ &\\
\mr
Pulsar2&$0.02$&$2.6 \cdot 10^{-13}$&$1.3 \cdot 10^5$&$17$&$2.2 \cdot 10^6$ &\\
Pulsar3&$0.04$&$3.0 \cdot 10^{-13}$&$2.6 \cdot 10^5$&$20$&$5.2 \cdot 10^6$&\\
Pulsar8&$0.20$&$2.2 \cdot 10^{-13}$&$1.3 \cdot 10^6$&$15$&$ 1.9  \cdot 10^7$&\\
\br
\end{tabular} 
\end{indented}
\end{table}

  %A map of the detection statistic obtained in the case of pulsar\_8 is shown in Fig. \ref{fig:p8_color_simo_zoom}.
%\begin{figure}[h!]
%\hspace*{-0.8cm}
%\includegraphics[scale=0.3]{immagini/p8_gray.jpg}
%\caption{Computed values of the detecton statistic over the parameter space explored for the HI pulsar\_ 8. The $\delta$-like and strongest peak is the nearest to the signal frequency and spin-down value. The color bar shows the logarithmic value of the detection statistic. The black circle shows the loudest candidate.}
%\label{fig:p8_color_simo_zoom}
%\end{figure}
%The loudest candidate, indicated by the circle,  is in the grid point nearest to the injected signal.
 From the loudest candidate we have estimated the signal parameters, obtaining values in good agreement with the injected ones. In fact, they are the best estimation we can obtain with the given search grid. In Tab. \ref{tabl3} we report the test results for the three considered hardware injections. Note, in particular, that the frequency error is always smaller than 1 bin, while the error in the estimation of the spin-down is always zero because the grid we have used for this test was centered at the exact spin-down of each signal.
\Table{\label{tabl3} Recovered parameters for the hardware injection analysis: $\epsilon_f$ is the mismatch, in units of frequency bin, between the injected
frequency and the detected frequency; $\epsilon_{\dot{f}}$ is the the mismatch between the injected spin-down value and the detected spin-down value; $\epsilon_{H_0}$ is the ratio of the estimated and injected amplitude; $\epsilon_\eta$ the normalized relative error on the parameter $\eta$ and $\epsilon_\psi$ the normalized relative error on the parameter $\psi$.}
\br
&\centre{5}{Relative errors}\\
&\crule{6}\\
Name&$\epsilon_f$&$\epsilon_{\dot{f}}$ &$\epsilon_{H_0}$ &$\epsilon_\eta$ &$\epsilon_\psi$ &\\
\mr
Pulsar2&$0.0373$&$0$&$1.041$&$0.0307$&/$^{\rm a}$&\\
Pulsar3&$0.1835$&$0$&$1.100$&$0.01$&$0.0085$&\\
Pulsar8&$0.0465$&$0$&$0.9833$&$0.0115$&$0.0225$&\\
\br
\end{tabular}
\item[]$^{\rm a}$ Pulsar\_2 is nearly circularly polarized and the polarization angle $\psi$ is then ill-defined and, consequently, badly estimated. 
\end{indented}
\end{table}

\subsection{Computational cost}
We have performed a narrow-band analysis for the Crab pulsar, similar to the one described in \cite{rob:obs} and compared the corresponding computational costs. The analysis in \cite{rob:obs} covered a frequency range $\Delta f=0.02 \, \textrm{Hz} $ and a spin-down range $\Delta \dot{f}=2.49 \cdot 10^{-13} \, \textrm{Hz}/s $.
For testing purposes we have repeated the narrow-band searches with the improved pipeline covering a larger frequency and spin-down range.\\
Specifically, we have used a frequency range of $\Delta f= 0.1 \,\textrm{Hz}$ and a spin-down range of $ \Delta \dot{f}=6.40 \cdot 10^{-13} \, \textrm{Hz}/s$ around the "central" values.
The corresponding number of points in the parameter space is given, for both the old and the new pipeline, in Tab. \ref{tabl4}.
\Table{\label{tabl4} Explored volume in the parameter space for the narrow-band searches performed for the Crab pulsar with the "old" and new pipelines. $\Delta f$ is the explored frequency range, $\Delta \dot{f}$ the explored spin-down range, $n_f$ and  $n_{\dot{f}}$ are the corresponding number of explored bins, and $n_{tot}$ is the total number of points in the parameter space.}
\br
Pipeline&$\Delta f$ [Hz]&$ \Delta \dot{f}$[Hz/s]&$n_{f}$&$n_{\dot{f}}$&$n_{tot}$\\
\mr
Old&$0.02$&$2.49 \cdot 10^{-13}$&$1.6 \cdot 10^{5}$&$33$&$5.8 \cdot 10^{6}$\\
New&$0.1$&$6.40 \cdot 10^{-13}$&$8.16 \cdot 10^{5}$&$165$&$1.34 \cdot 10^{8}$\\
\br
\end{tabular}
\end{indented}
\end{table}
For both the old and the new analysis we consider only one single value for the second order spin-down. This is due to the fact that the VSR4 run is short enough that the frequency variation due to second order spin-down is smaller than one frequency bin.
In the current test the number of points in the  parameter space is $n_{tot}^{new}/n_{tot}^{old} \approx 26 $ times bigger than in the old analysis. The total
computational time for the entire analysis, using the new pipeline, was of about 10 hours with a single job. The old analysis was split in 33 jobs, corresponding to the number of spin-down corrections, each long about 10 hours. The analyses have been carried on machines with similar characteristics. This means that the new method is nearly 3 order of magnitudes faster than the previous one at fixed parameters space. This huge speed increase makes the pipeline suited also for wider band searches, like those for neutron stars with known position and more uncertain rotational parameters. This are called {\it directed searches}. An example of such kind of search is discussed in the next section.

\section{Application to a directed search}
\label{sec:V}
Semi-coherent directed searches have been done in the past for the central compact object in Cassiopeia A \cite{abadie:casA} and recently for other few supernova remnants \cite{aasi:SNR}.
In order to show the possible extension to this kind of search of the pipeline described in this paper, we have performed a fully coherent directed search on Virgo VSR4 data for the central compact object in a supernova remnant, which rotational parameters are unknown. 

\subsection{Target description}
The central compact object (CCO) XMM UJ173203.3-344518 was discovered by XMM-Newton in 2007 \cite{tian:HESS} \cite{abra:hess} as a point-like X-ray source near the center of the TEV-emitting supernova remnant HESS J1731-347, a.k.a. G353.6-0.7.
The supernova remnant has a large angular size of $ \approx 0.5 \, \textrm{deg} $, from which a distance of 3.2 kpc and an estimated age of about 27 kyr  have
been deduced \cite{tian:HESS2010}. 
The celestial coordinates of the central compact object are $\alpha$=263.04166 deg and $\delta$= -34.7667 deg. No pulsation was observed hence we have no estimation of the rotational frequency and spin-down values of the neutron star. 
We can estimate the indirect spin-down limit for the amplitude of the gravitational waves using  Eq. \ref{eq:sdlimit}. For a distance of 3.2 kpc and an age of 27 kyr we obtain:
\begin{equation}
h_{0,in} \approx 1.4 \cdot 10^{-25}
\label{eq:gw_G353}
\end{equation} 
Considering the number of points in the parameter space, see next section, the estimated sensitivity of the search is \cite{rob:method} 
\begin{equation}
h_{sens}(f) \approx 25 \sqrt{\frac{S_{n}(f)}{T_{obs}}}
\label{eq:blabla}
\end{equation} 
Given the typical VSR4 noise level and duration, the sensitivity given by Eq.(\ref{eq:blabla}) is about 3 times larger than the indirect spin-down limit in the frequency range of the best detector sensitivity. 
Although this search has a limited scientific value, we have used it to show the feasibility of directed searches with the pipeline we have developed. 
\subsection{Selection of the parameter space}
We have chosen a 20 Hz frequency band from 116 Hz to 136 Hz, in which the noise level is relatively stationary. 
To define the spin-down range we have used the approximate relation which connects the frequency derivative to the frequency, the neutron star age $\tau$ and the braking index $n$:
\begin{equation}
\dot{f} \approx -\frac{f}{(n-1) \tau}
\end{equation}
Following the same approach used by \cite{wette:cassiopea} and \cite{abadie:casA}, we have considered a range of braking indexes between $n_{min}=3$ and $n_{max}=7$ which correspond, respectively, to magnetic dipole \cite{aasi:SNR} and {\it r-modes} emission \cite{owen:rmode}.
As a consequence, the spin-down range is given by 
\begin{equation}
-\frac{f}{(n_{min}-1) \tau} \leq \dot{f} \leq - \frac{f}{(n_{max}-1) \tau}
\label{eq:sd_range}
\end{equation}
For the second order spin-down we have used the definition of braking index, obtaining a range 
\begin{equation}
\frac{\dot{f}^2 n_{min}}{f} \leq \ddot{f} \leq \frac{\dot{f}^2 n_{max}}{f}
\label{eq:ssd_range}
\end{equation}
Given VSR4 run duration, $T_{obs}\simeq 8.14 \cdot 10^6 s$, the resulting bin widths are 
$\delta f=1.2285 \cdot 10^{-7} \, \textrm{Hz} \ $, $ \delta \dot{f}=1.509 \cdot 10^{-14} \textrm{Hz}/s$,
$ \delta \ddot{f}=3.70 \cdot 10^{-21} \textrm{Hz}/s^2$.
corresponding to the following number of points in the parameter space:
$n_f \approx 1.6 \cdot 10^{8},~n_{\dot{f}} \approx 3300,~n_{\ddot{f}} =1,~n_{tot} \approx 4.8 \cdot 10^{11}$.
For practical reasons the search band has been divided in ten 2 Hz sub-bands later down-sampled at 2 Hz. This part of the analysis took bout 18 hours on a 8-core CPU. 
The band 116-136 Hz is populated by several known disturbances. The polluted frequencies have not been taken into account in the computation of the detection statistic. This disturbances are listed in Tab. \ref{table5}. The very noisy band from 130 Hz to 132 Hz has been excluded from the analysis since the beginning.
\Table{\label{table5} Known noise lines in the frequency band 116-136 Hz of VSR4 data. These bands have not been considered in the computation of the detection statistic.}
\br
\textbf{Frequency}  [Hz] & \textbf{Width} [Hz] & \textbf{Origin} \\ 
\mr
116.5 &	0.015& Laser calibration line \\
120.006 &	0.0055 & Harmonic of 10 Hz \\
123.343 &	0.0085 & Harmonic of 10.2872 Hz \\
133.622 &	0.0085 & Harmonic of 10.2872 Hz \\
\br
\end{tabular}
\end{indented}
\end{table}
In Tab. \ref{table6} the main parameters of the search for each 2-Hz band are reported. The total observation time is not long enough to appreciate more than one second order spin-down bin.
\Table{\label{table6}Search parameter space for XMM UJ173203.3-344518. The number of frequency bins explored in each 2 Hz sub-band is $n_{f}=1.63 \cdot 10^7$.}
\br
\centre{1}{\textbf{Frequency}} &&& \centre{1}{\textbf{Spin-down}}  &&& \textbf{$n_{\dot{f}}$} && \centre{1}{$\ddot{f} $}  \\
\textbf{band} [Hz] &&&\textbf{band} [Hz/s $\cdot 10^{-11} $] &&& && [Hz$/s^2  \cdot 10^{-23}$] \\
\mr
116-118  &&& $[-6.9246;-2.2691]$  &&& 3083 && \centre{1}{ $9.0464$} \\
118-120 &&& $[-7.0419;-2.3082]$  &&& 3134 && \centre{1}{$8.8956$} \\
120-122 &&& $[-7.1593;-2.3473]$  &&& 3186 && \centre{1}{$8.7498$} \\
122-124 &&& $[-7.2767;-2.3864]$  &&& 3238 && \centre{1}{$8.6086$} \\
124-126 &&& $[-7.3940;-2.4256]$  &&& 3290 && \centre{1}{$8.4720$} \\
126-128 &&& $[-7.5414;-2.4647]$  &&& 3342 && \centre{1}{$8.3396$} \\
128-130 &&& $[-7.6288;-2.5429]$  &&& 3394 && \centre{1}{$8.2113$} \\
130-132 &&& \centre{1}{-}  &&& \centre{1}{-} && \centre{1}{-} \\
132-134 &&& $[-7.7461;-2.5429]$  &&& 3497 && \centre{1}{$7.9662$} \\ 
134-136 &&& $[-7.9800;-2.6212]$  &&& 3549 && \centre{1}{$7.8490$} \\
\br 
\end{tabular}
\end{indented}
\end{table}\newline
The corresponding value of the frequency second derivative has been obtained, for each band of 2 Hz, taking the mean value of $\ddot{f}$ predicted from the braking index relation using breaking indexes n=3,5,7.  The total explored parameter space  consists of $n_{tot} \simeq 4.8 \cdot 10^{11} $ points.
The detection statistic has been computed for each point in the parameter space and we have taken the local maxima of the detection statistic every 0.001 Hz and over the full spin-down range. Keeping into account the "look-elsewhere" effect, as was done in \cite{rob:obs}, we have selected as candidates those local maxima in the frequency/spin-down plane with a value of the detection statistic corresponding to an overall p-value of 1\% or less on the noise only distribution.

\subsection{Candidates}
\begin{figure}[h!]
%\hspace*{-3.5cm}
\includegraphics[width=1.\textwidth]{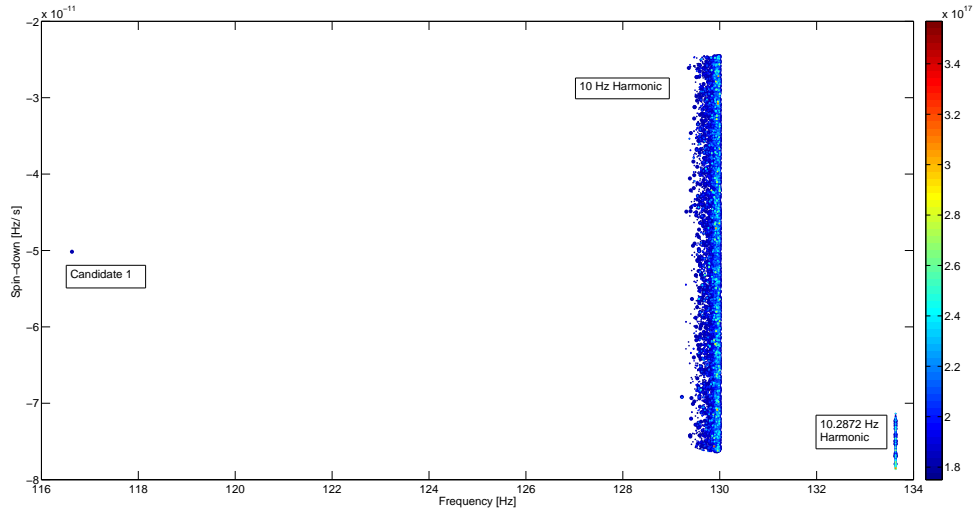}
\caption{Candidates with a value of the detection statistic over the threshold corresponding to a p-value equal to 1\% .}
\label{fig:candidate}
\end{figure}

\begin{figure}
\centering
\subfloat[][Data power spectrum around Candidate 1. The green dashed line corresponds to the frequency of the candidate 1. \label{fig:candidate1_out}]
{\includegraphics[width=.5\textwidth]{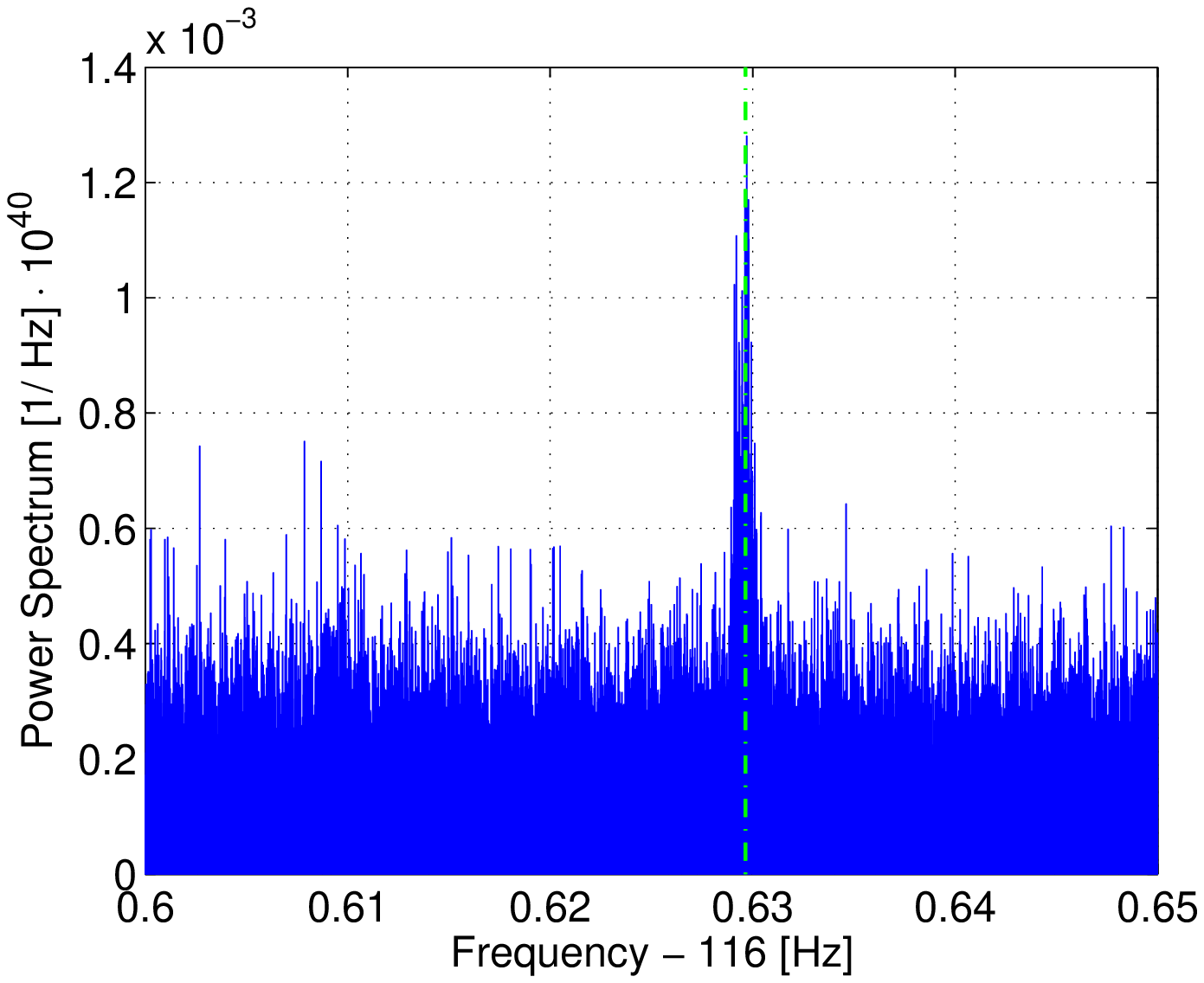}} \qquad 
\subfloat[][Zoom of the power spectrum in Fig. \ref{fig:candidate1_out} around Candidate 1. The  green dashed vertical lines are the expected frequencies of a
CW signal associated with Candidate 1. \label{fig:candidate1}]
{\includegraphics[width=.5\textwidth]{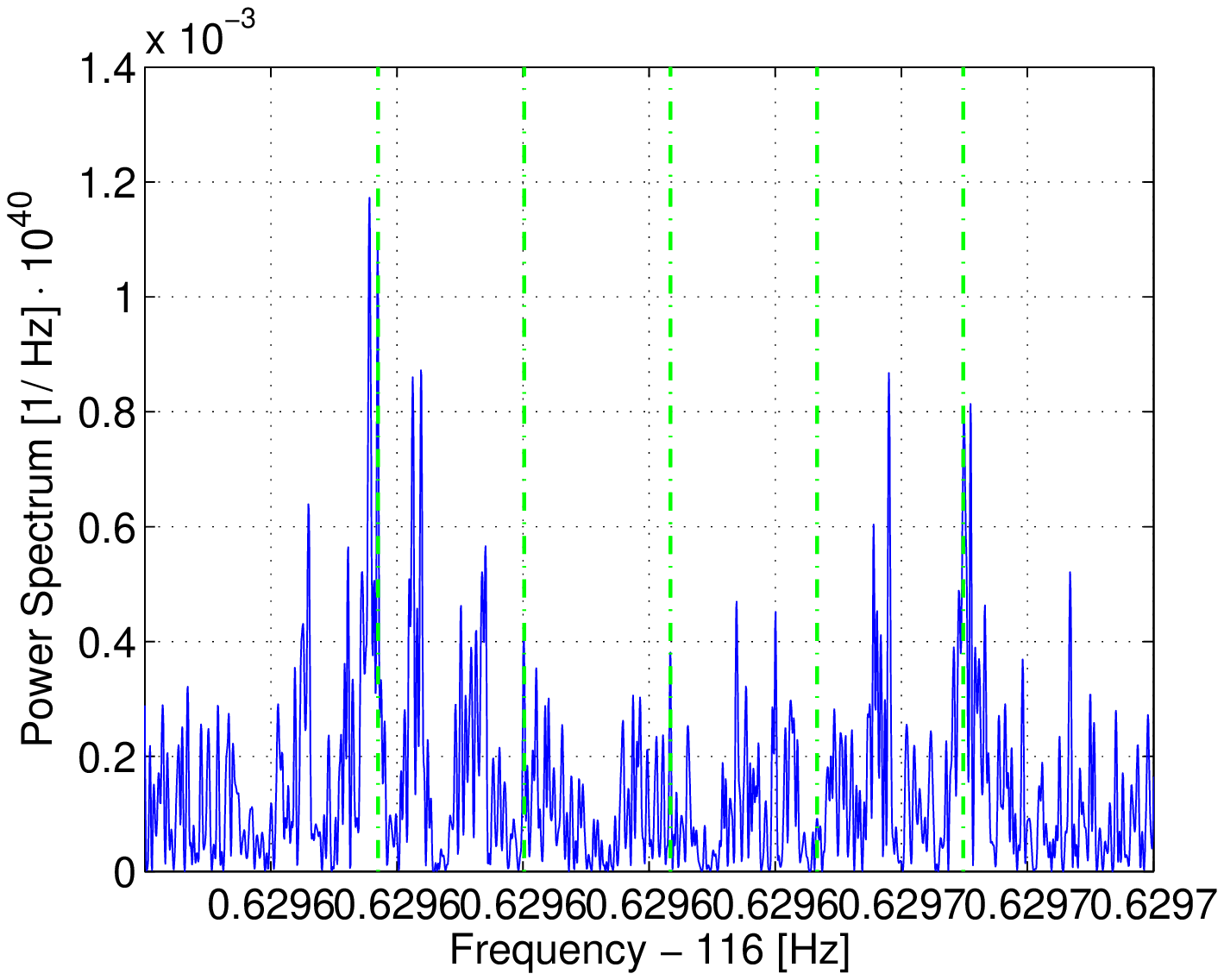}}
\end{figure}

The selected candidates are plotted in Fig. \ref{fig:candidate}.
They are grouped in three different sets. 
Two of them are due to instrumental disturbances, although they are not in the official noise lines list. In particular, the set of candidates near 130 Hz has
been identified by the noise lines hunting tool NoEMi\footnote{Noise Frequency Event Miner (NoEMi) is a tool able to indetify on a daily basis, the frequency
lines observed in the Virgo science data looking for coincident lines present in different environment sensors \cite{accadia:NOEMI} \cite{accadia:NOEMI2}} as
due to an harmonic of the 10 Hz comb, the same that generated the 120.006 Hz line listed in Tab. \ref{table5}. These harmonics have several side-bands
extending in a range of $\sim 0.4 Hz$ around the central frequency, which are the likely source of this set of candidates.
The set of candidates at 133.648 Hz has been associated by NoEMi to an harmonic of the 10.2872 Hz comb, also shown in Tab. \ref{table5}. Although the nominal frequencies corresponding to this line have been removed before the computation of the detection statistic, the application of the Doppler correction shifts the line frequency pushing some candidates outside the vetoed band. There is just one remaining candidate, indicated as \textbf{Candidate 1} in Fig. \ref{fig:candidate}, which cannot be associated to any known instrumental disturbance. This candidate has the following parameters:
\begin{eqnarray}
& f= 116.6296417  \, \textrm{Hz} \qquad \dot{f}=-5.01608 \cdot 10^{-11} \,  \textrm{Hz}/s \\
& \qquad \qquad \qquad \ddot{f}= 9.20232 \cdot 10^{-23} \, \textrm{Hz}/s^2 
\label{eq:param}
\end{eqnarray}
The overall associated  p-value is 0.006 which, in the case of noise with Gaussian distribution, would correspond to $\sim$2.5$\sigma$. No known noise line was found neither by 
NoEMi 
neither in the noise spectra at the candidate's frequency. However, this is not an outstanding candidate, and could be simply due to a noise fluctuation. Moreover, a detection at this level of sensitivity would require some non trivial explanation for why the signal strain amplitude is above the spin-down limit. Nevertheless, we have decided to proceed with some basic follow-up steps in order to confirm or reject the candidate. In Fig. \ref{fig:candidate1_out} the data power spectrum around the candidate is plotted,showing a region of excess power around the candidate frequency. In Fig. \ref{fig:candidate1} a zoom of the power spectrum is shown. The five dashed vertical lines correspond to the frequencies at which the signal power would be split due to the sidereal motion of the detector. There are indeed peaks at frequencies near the expected ones, but no clear correspondence. 
As a further check, we have performed a targeted search for Candidate 1 in the data of the second science run of Virgo (VSR2). VSR2 extended from July-7th-2009 21h UTC to January-8th-2010, 22h UTC, two years before VSR4. Considering that VSR2 run has a sensitivity comparable to that of VSR4 in the explored frequency region, if \textbf{Candidate 1} arises form a continuous gravitational wave signal then it should be detectable also in this data set. The rotational parameters of the candidate given in Eq. \ref{eq:param}, which refer to the starting time of VSR4, have been shifted to the VSR2 starting epoch by taking into account the spin-down,  obtaining the following parameters:
\begin{eqnarray*}
& f= 116.63265826 \textrm{Hz}  \dot{f}=-5.0166 \cdot 10^{-11} \textrm{Hz}/s \\
& \qquad \qquad \quad \ddot{f}= 9.20232 \cdot 10^{-23} \textrm{Hz}/s^2
\end{eqnarray*}
A full coherent targeted search has been done using these parameters and the power spectrum of the corrected time series has been computed to look for the presence of  significant peaks.
\begin{figure}
\centering
\subfloat[][Power spectrum of VSR2 data around the frequency of Candidate 1 after having performed a fully coherent search targeted at Candidate 1's parameters. No significant peak is visible. \label{fig:candidateVSR2}]
{\includegraphics[width=.5\textwidth]{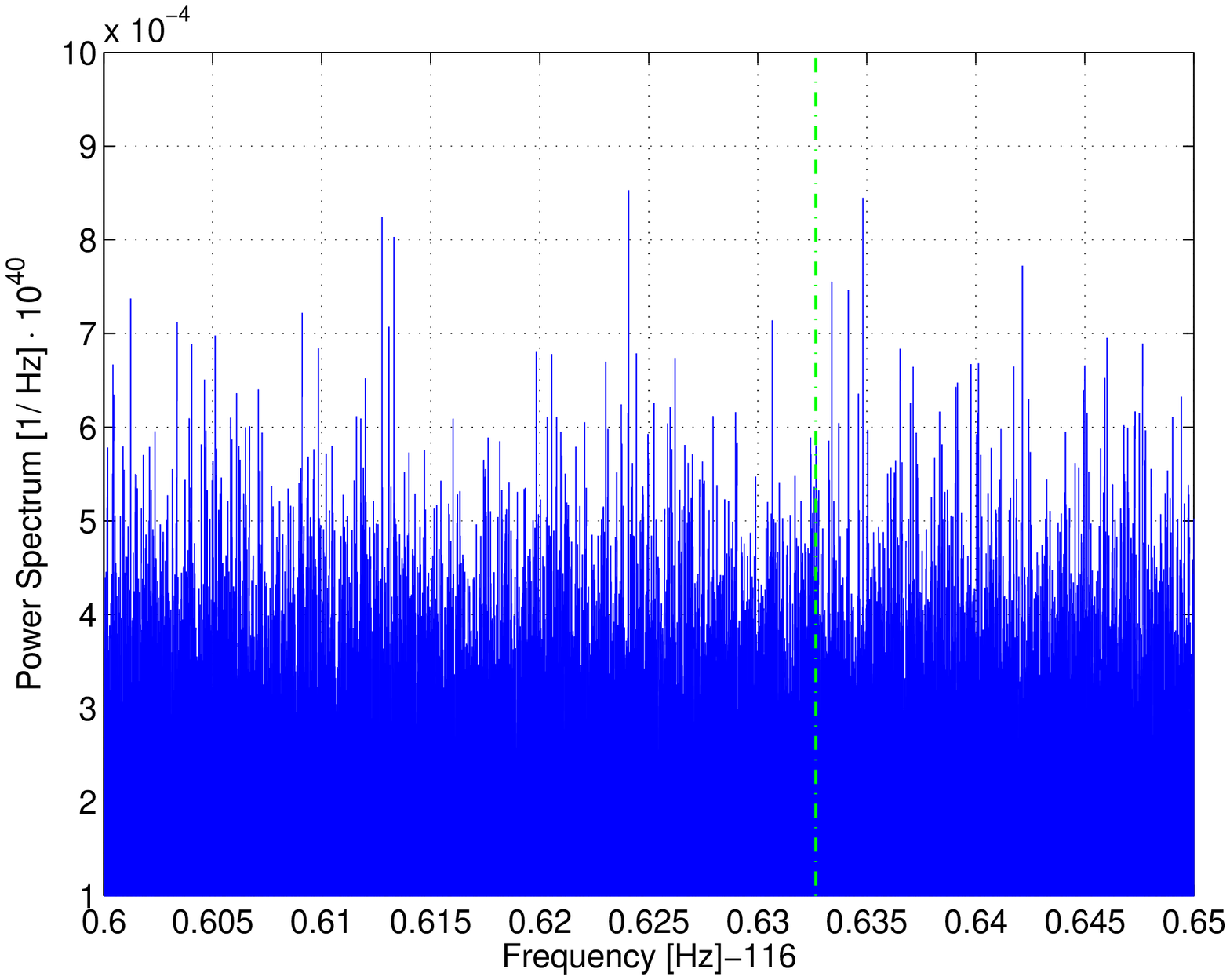}} \qquad
\subfloat[][Zoom of the power spectrum in Fig. \ref{fig:candidateVSR2} around Candidate 1 in VSR2 data. The  green dashed vertical lines are the expected frequencies of a signal associated with Candidate 1. \label{fig:candidateVSR2_faraway}]
{\includegraphics[width=.5\textwidth]{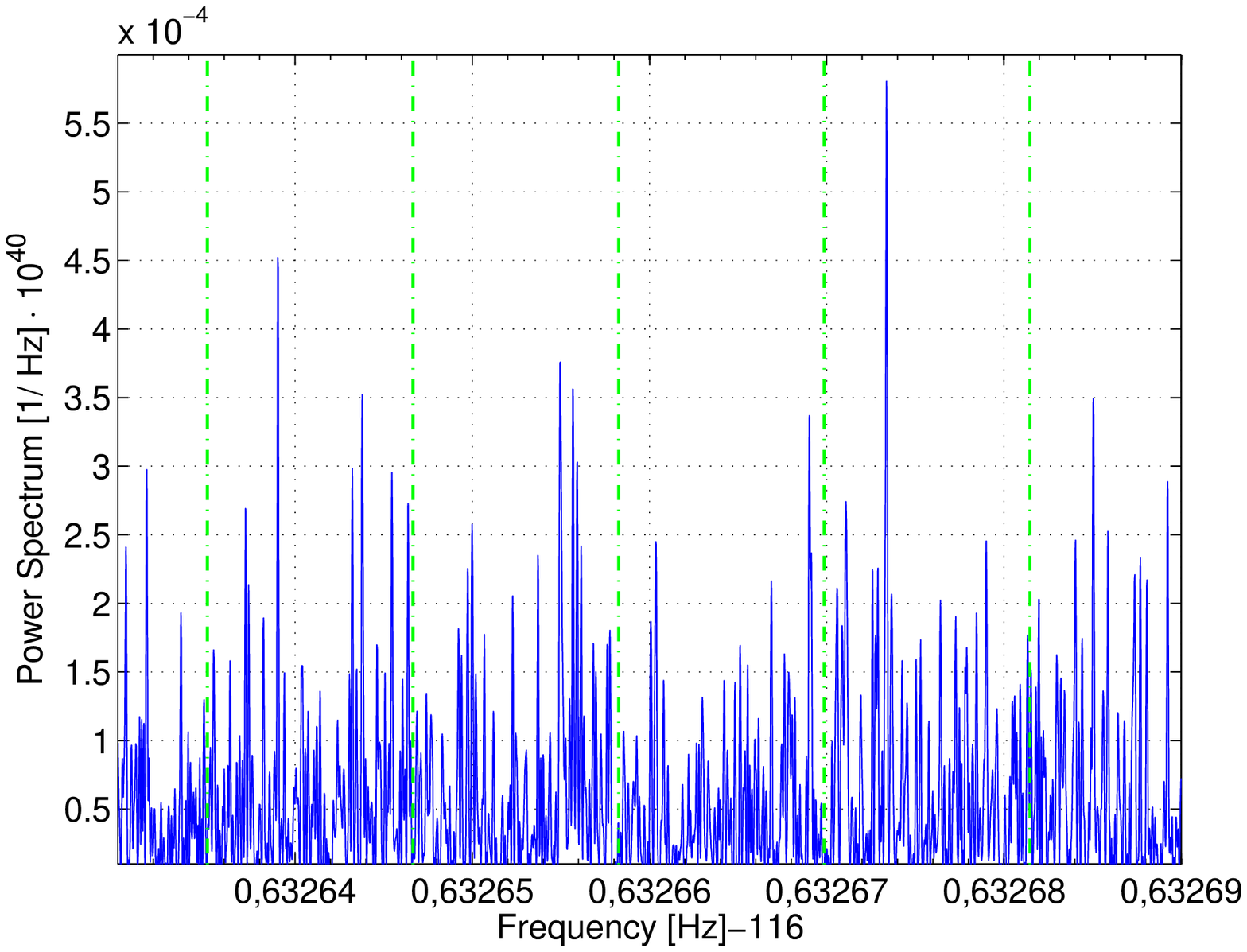}}
\end{figure}
The power spectrum is shown in Fig. \ref{fig:candidateVSR2}. The data appear quite clean and no significant peak is visible. The corresponding value of the detection statistic is fully compatible with noise. We conclude it is very unlikely \textbf{Candidate 1} is due to a GW signal. 
\newpage
\section{Conclusions}
\label{sec:VI}
The search for continuous graviational wave signal emitted by known spinning neutron stars, such as pulsars, typically assumes that the gravitational wave
frequency is at a known fixed ratio with respect to the star rotational frequency, thus implying that the gravitational signal is "phase locked" to the
electromagnetic one. Fully coherent searches, based on matched filtering, in principle offer the best sensitivity. However, they assume signal parameters are
known with high accuracy. Due to our ignorance of the mechanisms at the base of both the electromagnetic and gravitational emission in spinning neutron stars,
we cannot exclude a small mismatch between the gravitational wave frequency and that inferred from electromagnetic observations. For long observation times,
even a very small discrepancy can prevent signal detection. In order to take into account such possible mismatch, it is very important to perform narrow-band
searches with the best possible sensitivity, that, in layman's terms means to apply matched filtering on several points of the source parameter space. The computational load of the analysis rapidly increases with the searched volume and the observation time. For these reasons past narrow-band searches were limited to frequency bands of a few milliHertz, and to few spin-down values around the central search values inferred from EM observation \cite{rob:obs}.

In this paper we have presented an improved pipeline for narrow-band searches of continuous gravitational wave signals emitted by asymmetric spinning neutron stars. The new pipeline is about three orders of magnitude faster than the old version allowing to perform, in few hours of CPU time on a standard workstation, a full coherent search of months of data over a frequency band of $\mathcal{O}(1$ Hz$)$ and several hundreds of spin-down values. 
In the future, the use of many-core processors, like GPU, could boost the analysis further. Currently, this novel pipeline allowed us to make a narrow-band
search on Advanced LIGO O1 data for 11 known pulsars for some of which no updated ephemeris were available. The results of this analysis will be described
elsewhere \cite{simone:O1_narrow}. 
We have also shown the possibility to extend this algorithm to directed searches of sources, like central compact objects in supernova remnants such as G353.6-0.7, for which the position is accurately known while the rotational parameters are uncertain. Despite with VSR4 data it was not possible to beat the spin-down limit, it will become possible using the advanced detectors data such as the first obervational run of LIGO (O1). Thus making narrow-band searches very important to study this type of astrophysical sources. 

%\newpage

\end{document}